\documentstyle[prl,aps,epsfig,twocolumn]{revtex}

\begin{document}

\draft
\twocolumn[\hsize\textwidth\columnwidth\hsize\csname @twocolumnfalse\endcsname
\title{Fluctuation-dissipation relation in a sheared fluid}
\author{Jean-Louis BARRAT$^{1}$ and Ludovic BERTHIER$^{1,2}$}

\address{$^1$D\'epartement de Physique des Mat\'eriaux,
Universit\'e C. Bernard and CNRS, 
F-69622 Villeurbanne Cedex, France}

\address{$^2$Laboratoire de Physique, ENS-Lyon and CNRS,
 F-69007 Lyon, France}

\date{\today}

\maketitle

\begin{abstract}
In a fluid out of equilibrium, the fluctuation dissipation theorem (FDT)
is usually violated. Using molecular dynamics simulations, we study in detail
the relationship between correlation and
response functions in a fluid  driven into a stationary non-equilibrium state.
Both the high temperature fluid state and the low temperature glassy state
are investigated. In the glassy state, the violation of the FDT
is similar to the one observed previously in an aging
system in the absence of external drive. In the fluid state, violations of the
 FDT  appear only when the fluid is driven beyond the linear response regime, 
and are then similar to those observed in the glassy state.  
These results are consistent with the  picture obtained earlier
 from theoretical studies of driven mean-field disordered models, confirming 
 the  similarity  between these models and simple glassy systems.  
\end{abstract}

\pacs{PACS numbers: 05.70.Ln, 64.70.Pf, 83.50.Gd \hspace*{4.3cm} 
LPENSL-TH-03/2000}


\vskip2pc]

\narrowtext

In the past years, a large body of theoretical 
\cite{aging_th}, experimental \cite{aging_exp} and
numerical \cite{aging_num,fdt_lj} work has been devoted 
to the study of correlation
 and response functions in non-equilibrium glassy systems.
Generally, the non-equilibrium situation of interest 
is generated by a quench below the glass transition temperature.
The system falls then out of equilibrium, in the sense that its relaxation
time becomes far greater than the experimental time scale. 
Its properties therefore depend
on the time $t_w$ elapsed after the quench, also called  waiting time. 
The most spectacular dependency is observed for the two-time 
correlation functions which depend then upon the two time arguments and not 
on the time difference only, as in an equilibrium system. 
These correlations   decay with a relaxation
time $t_r$ that increases with $t_w$ (aging), typically $t_r \propto t_w$. 
The same aging behavior is observed in the response functions, and
a useful quantification of the non-equilibrium
behavior is encoded in the way the usual 
equilibrium fluctuation dissipation theorem (FDT)
is violated \cite{leticia3}.
For atomic systems of the 
type studied in this paper, this violation has been shown by numerical 
studies to be similar to the one observed in high dimensional
disordered mean-field models \cite{fdt_lj}.

These studies, however, focus on the case 
of a non-equilibrium situation following a quench and 
on the subsequent aging
phenomena.
In Ref. \cite{leticia3}, 
Cugliandolo {\it et al} suggested another approach
to non-equilibrium systems, in which the non-equilibrium state
was generated by ``stirring'' the system.  
In such a situation,
in which energy is fed into the system at a constant rate, the theory 
predicts that
a stationary non-equilibrium state is reached,
even when the unstirred system is in a glassy 
state: {\it aging is stopped} \cite{leticia3,ledou}. 
Further theoretical
studies on driven mean-field disordered systems \cite{BBK}
established a detailed picture of the non-equilibrium 
behavior, which can be  summarized as follows.
The relaxation of these systems is a two-step process composed of a fast
part which is essentially unaffected by the driving force, and a slow
relaxation occurring on a time scale which is a decreasing function
of the drive intensity.
Simultaneously, a two-temperature pattern appears, the fast
modes being equilibrated at the bath temperature $T$, while the slow ones 
have an effective temperature $T_{\text{eff}} > T$.
Quantitatively, this is shown by studying  a
correlation function $C(t) = \langle A(t_0+t) B(t_0) \rangle $ 
between two observables $A$ and $B$, 
and the associated response function $\chi(t)= \delta \langle A(t_0+t)
\rangle / \delta h(t_0)$,
where $h$ is the field conjugated to $B$. 
Whereas
at equilibrium, these quantities are related by the FDT, 
$\partial_t C(t)= -  k_B T \chi(t)$, 
this FDT has to be generalized in the driven system by 
introducing an effective temperature 
$T_{\text{eff}}$, through $\partial_t C(t) = -  k_B 
T_{\text{eff}}\chi(t)$.
In the limit of zero drive,  one finds $T_{\text{eff}} \rightarrow T$ 
if $T>T_c$,
($T_c$ is the temperature 
at which the relaxation time of the undriven system diverges) and
 equilibrium properties are recovered.
For $T<T_c$, the limiting effective temperature coincides with 
that of the system {\it aging} at the same temperature \cite{leticia3,BBK}.
In Ref. \cite{franz} it was shown that,  for mean field
models, $T_{\text{eff}}$ is  
related to the configurational entropy available
to the system near one free energy minimum, confirming the 
interpretation of $T_{\text{eff}}$ as a true 
temperature \cite{leticia3,theo}, both in the 
thermodynamic and dynamical sense.

In this paper, we investigate the non-equilibrium situation
created by a steady, homogeneous  shear 
imposed on a simple glass forming liquid, {\it i.e.} a simple
realization of the stirring systems considered in Ref.\cite{leticia3}
which was suggested by Ref.\cite{BBK}.
Note however that, in practice, such a situation is  more easily 
realized  in a ``soft'' glassy system (a complex fluid with glassy 
behavior \cite{peter}), which can support homogeneous shear flow,
than in a usual molecular glass, in which shear banding and fracture 
tend to take place. Simulations, on the other hand,  allow to create an
 homogeneous shear even in a simple system. 
The goal of this paper is then to determine the behavior of 
a simple glassy system subjected to a shear and to compare the results with
the main predictions of mean-field calculations, keeping in mind
the possible relevance to soft glassy materials. Although the mean-field 
scenario is not a priori expected to apply to a three dimensional fluid,
it has the advantage of providing precise predictions, which can be compared
easily to experimental or numerical results. 
Our focus will be on the 
fluctuation-dissipation relation, the general rheological properties 
already discussed by Yamamoto and Onuki \cite{Onuki}  will be
only briefly considered.

The system simulated in this work is a 80:20 mixture 
of 2916 Lennard-Jones particles,
with interaction parameters that prevent crystallization~\cite{kob}.
In all the paper, the  length, energy and time units are the 
standard Lennard-Jones units $\sigma_{AA}$ (particle diameter),
$\epsilon_{AA}$ (interaction energy), and $\tau_0 = (m_A\sigma_{AA}^2/
\epsilon_{AA})^{1/2}$ \cite{note1}, where $m_A$ is the particle mass and the
subscript $A$ refers to the majority species~\cite{kob}. The system has
been described in detail elsewhere, and its equilibrium (high
temperature) properties have been fully characterized. At the reduced
density $\rho=1.2$, where all our simulations are carried out, 
a ``computer glass transition'' is found in the
vicinity of $T_c=0.435$ and the slowing down of the dynamics seems to be
described well by Mode-Coupling theory~\cite{kob}. The aging behavior
of the system below this temperature has also been characterized extensively,
including the violation of the FDT in the glassy phase \cite{barratkob}.

The homogeneous shear state corresponding to a planar Couette 
flow is obtained by using the SLLOD algorithm
supplemented by Lees-Edwards boundary conditions \cite{AT87}.
The velocity gradient is in the $z$ direction, and the fluid velocity in the
$x$ direction.
Constant temperature conditions are ensured by thermostatting the 
 velocities in the direction perpendicular to the 
flow using the   Nos\'e-Hoover method \cite{AT87}, or in some cases through
a simple 
 velocity rescaling.
The shear rate, denoted by ${\dot\gamma}$, 
naturally introduces a new time scale 
$\dot\gamma ^{-1}$ into the problem. Obviously, 
 a simulation involving a steady shear state is possible only if
the available simulation time is significantly larger 
than $\dot\gamma ^{-1}$. 
This limits our study to shear rates larger than typically $10^{-4}\tau_0$,
corresponding to $10^6$ time steps. 

The first important consequence of applying a shear to the system is that,
in accordance with theoretical expectations, a 
non-equilibrium stationary state
is reached after  a transient of a few $\dot\gamma ^{-1}$. 
This is true even at temperatures at which
the corresponding undriven system 
behaves like a glass and does not reach equilibrium. 
The benefit is that many of the difficulties associated with the simulation 
of glassy systems, such as the dependence on the preparation method
of the sample (especially cooling rate), are eliminated. At a given 
temperature, a sheared sample can be prepared either by cooling another 
sheared sample from a higher temperature,
 or by shearing a  sample quenched at zero shear. The results 
will be identical provided the first steps
 of the simulation are discarded. 
That the time translation invariance
property is recovered under shear is illustrated in Fig. 1, which displays
the  incoherent scattering function 
of a quenched and sheared sample
for several different waiting times after the quench. As usual, 
the incoherent scattering function for the particles of type $A$ and 
 wave-vector ${\bf q}$ is defined by
\begin{equation}
C_{\bf q}(t)= \frac{1}{N_A} \sum_{j=1}^{N_A}
\left\langle \exp\left(
i{\bf q}\cdot [ {\bf r}_j(t+t_0)-{\bf r}_j(t_0)] \right)  \right\rangle.
\label{fsqt}
\end{equation}
In this study, only values of ${\bf q}$ in the $y$ direction
(i.e. perpendicular to both  the velocity and  the velocity 
gradient) will be considered. This choice avoids the  complications 
due to convection by the average flow that arise for other wave-vectors
\cite{Onuki}.

\begin{figure}
\begin{center}
\psfig{file=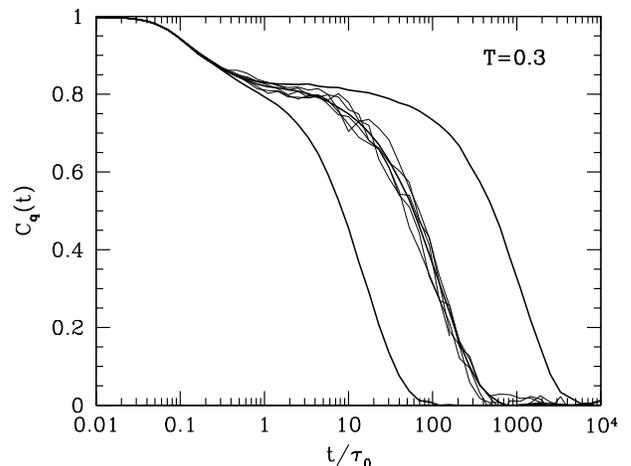,width=8.2cm,height=6.3cm}
\caption{Incoherent scattering functions for the A particles at 
$T=0.3$ and ${\bf q} =7.47 {\bf e_y}$, 
the location of the main peak in the structure factor.
 Bold curves, from
 left to right:
$\dot{\gamma}=0.01$, 0.001 and
0.0001. For $\dot{\gamma}=0.001$ 
five different two-time correlation functions,
taken with time origins
equally spaced during a run of duration $10^4\tau_0$
are also shown as light continuous curves.
 The absence of aging is illustrated  by the fact that these
curves coincide with the bold one.}
\label{fig1}
\end{center}
\end{figure}

\begin{figure}
\begin{center}
\psfig{file=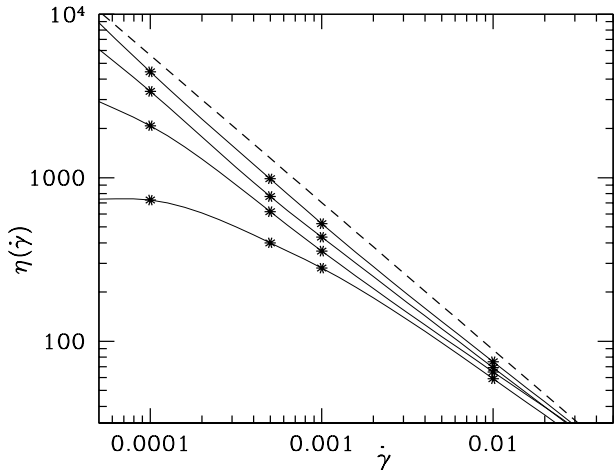,width=8.2cm,height=6.3cm}
\caption{Viscosity versus shear rate for temperatures 
(from bottom to top) 
$T=0.5$, 0.45, 0.4 and 0.3.
The solid lines are guides to the eye, the dashed
line corresponds to  a power law $\eta \propto \dot\gamma^{-0.9}$.}
\label{fig2}
\end{center}
\end{figure}

In Fig. \ref{fig1}, it is clearly seen that the relaxation time $t_r$ 
of the correlations is shear rate dependent.
As already shown in Ref. \cite{Onuki}, the viscosity $\eta$ scales
roughly as $t_r$.
Consequently, $\eta$ decreases when $\dot\gamma$ increases:
this is a {\it shear-thinning} behavior.
The viscosity is
defined by $\eta({\dot\gamma}) \equiv
\sigma_{xz}({\dot\gamma})/{\dot\gamma}$, where $\sigma_{xz}$ is the 
off-diagonal component of the stress tensor, and is shown in Fig. \ref{fig2}.
The same type of  rheological behavior was obtained in Ref. \cite{Onuki} for
a similar system. 
A Newtonian regime, where $\eta$ is independent 
of ${\dot\gamma}$, is obtained when
$\dot\gamma ^{-1} \gtrsim t_r$.
For $T<0.45$, no such regime is observable. 
The shear-thinning behavior is well characterized by a power law
$\eta({\dot\gamma}) \sim
{\dot\gamma}^{-\alpha(T)}$. 
For the lowest temperature investigated here,
$T=0.3$, one finds  $\alpha \simeq 0.9$. 
This shear-thinning exponent is slightly  smaller than
the one obtained in \cite{Onuki}. 
There is however no reason to expect a 
universal value for $\alpha$, which experimentally is found to be 
system dependent, with reported values between 
0.5 and 1 \cite{shear_thinning}.

The fact that the time translation invariance is reestablished by shearing
the system implies a superficial resemblance with equilibrium systems. The
correlation function displayed in Fig. \ref{fig1}, for example, 
is very similar to the correlation functions in a fluid system slightly above
the glass transition. In the following, the differences between the dynamics
of fluctuations in the driven and equilibrium system is studied through
the fluctuation-dissipation relation introduced in the introduction.
In the present case, the two observables under study are
$A_{\bf q}(t)=1/N_A \sum_j \epsilon_j \exp [ i {\bf q} \cdot {\bf r}_j (t) ]$, 
and $B_{\bf q}(t) = \sum_j \epsilon_j \cos [ {\bf q} \cdot {\bf r}_j (t) ]$,
where $\epsilon_j$ is a random variable taken from a
bimodal distribution $\epsilon_j = \pm 1$.
It is straightforward to show that after averaging over the distribution
$\{ \epsilon_j \}$, the correlation function 
$\langle A_{\bf q}(t+t_0) B_{\bf q}(t_0) \rangle$ is equal to 
$C_{\bf q}(t)/2$, where $C_{\bf q}$ is the incoherent scattering 
function defined in Eq. (\ref{fsqt}).
To compute the response function, a term $\Delta H = - h B_{\bf q}(t)$ 
is added to the Hamiltonian.
The response function is then $\chi_{\bf q}(t)= \delta \langle 
A_{\bf q}(t+t_0) \rangle / \delta h(t_0)$.
The procedure to study the FDT is then the following.
The system is made stationary at a fixed
temperature and shear rate.
The field is switched on at $t_0$ and
the observable
$A_{\bf q}(t + t_0)$ is monitored. 
The same procedure is repeated for
several (20 to 80) realizations of the charge distribution.
This gives the integrated response function $M_{\bf q}(t)$, defined as:
\begin{equation}
M_{\bf q} (t) = \int_{t_0}^{t+t_0} dt' \chi_{\bf q}(t') \simeq 
\frac{ \langle A_{\bf q}(t) \rangle}{h}.
\end{equation}
The last equality holds in the linear response regime, that we 
carefully checked, by choosing a small amplitude for the field $h$ (between
0.05 and 0.2).
This procedure was carried out for four different values of the 
temperature ($T=0.5$, 0.45, 0.4 and 0.3) both above and below the
computer glass transition temperature.  
At each temperature, shear rates
from $5. 10^{-5}$ to $10^{-2}$ were considered. 
The wave vector was ${\bf q} = 7.47 {\bf e_y}$. 
The results are most easily analyzed by considering parametric plots of
$M_{\bf q}(t)$ versus $C_{\bf q}(t)$. 
The slope of these curves is, by definition,  $-k_B T_{\text{eff}}$.

\begin{figure}
\begin{center}
\psfig{file=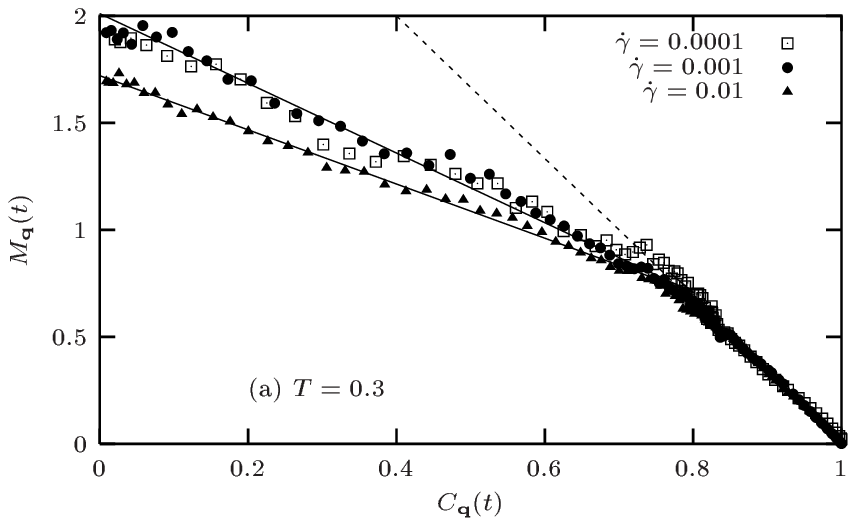,width=8.5cm,height=6.5cm}
\psfig{file=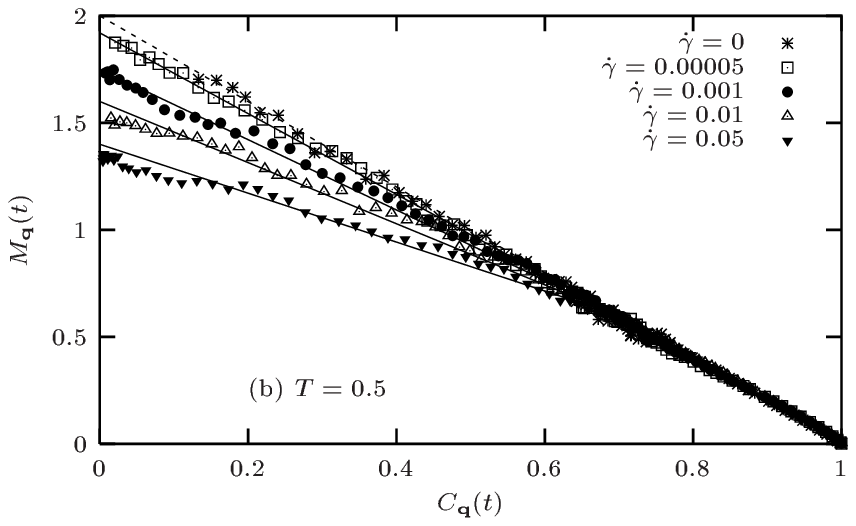,width=8.5cm,height=6.5cm}
\caption{Parametric plots for (a) $T=0.3$ and (b) $T=0.5$, and various 
shear rates. In both figures, the dashed line is the FDT, and has a slope
$-1/T$.
The full lines are linear fits to the data for $\dot{\gamma} > 0$.}
\label{fig3}
\end{center}
\end{figure}

In Fig. \ref{fig3}, such  plots are shown for (a) $T=0.3$, which 
is deep in the glassy region, and (b)
$T=0.5$, where the system
is at equilibrium in the absence of shear.
For $T=0.3$, a very clear 
deviation from FDT is observed when the correlation falls below
$C\simeq 0.8 =q$, which corresponds to the plateau value in the correlation
functions (see Fig. \ref{fig1}). The parametric curve can be very well
 approximated by
two straight lines, one with slope $-1/k_B T$ (FDT) at large correlations and
one with a slope $-1/k_B T_{\text{eff}}$ for $C< q$.
The latter slope saturates for the smallest shear rates to the value
$T_{\text{eff}} \simeq 0.62 > T_c$ (this straight line 
correctly fits the two values $\dot{\gamma}=0.001$ and 0.0001). 
Remarkably, this value for $T_{\text{eff}}$
is very close (certainly within error bars)
to the one obtained in Ref. \cite{barratkob},
when studying the aging system. 
Such a coincidence was
expected on theoretical grounds \cite{leticia3,BBK}, 
and exemplifies the deep 
meaning of the effective temperature. 
As it is in fact a property
of the free energy landscape available to the system, it naturally arises 
in both types of dynamics (aging and driven) of the system. 
The two-temperature pattern is also consistent with earlier expectations
that a simple atomic  glass has the same dynamical behavior
of mean-field disordered models of the ``$p$-spin'' type ($p>2$)
 \cite{aging_th,BBK}.
It is worth noting that 
the value $q$ below which the  violation of FDT
is observed is, again as expected from theory, approximately equal
to the plateau value of the correlation functions. 
This had not been found in Ref. \cite{barratkob}, and the present results
show that this was probably due
to preasymptotic effects in the aging results.

At $T=0.5$, the FDT holds at zero shear (this is very well
verified numerically, see Fig. \ref{fig3}a) 
and remains valid for small enough shear rates. 
Deviations are observed 
only for shear rates large enough
to induce non-Newtonian behavior and shear-thinning. 
In this regime, it is natural to expect that deviations to FDT 
will become more important for increasing shear rates.
The data clearly indicate that 
the parametric plot {\it can
still be fitted by two straight lines}, with a
shear rate dependent slope for the non-FDT part. 
If this slope is interpreted as an effective temperature,
this temperature increases with increasing the strength of the driving force,
which is a quite intuitive effect.
Again, this is very similar to the results obtained in \cite{BBK} for
the disordered mean-field model.
Finally, we note that the results obtained for $T=0.4$ and $T=0.45$ are 
similar to those obtained for $T=0.3$ and $T=0.5$, respectively. 

In this paper, we have presented the first numerical study
of the fluctuation-dissipation relation in a realistic model
of a sheared fluid. 
This study demonstrates that the non-equilibrium
fluctuations obey the {\it two-time scales, two-temperature scenario}
which was previously derived for mean-field glass models \cite{BBK}.
Although the system is a three dimensional one, which will
certainly differ from mean-field systems in many respects, the 
mean-field approach provides here a useful conceptual framework
to rationalize the numerical findings.
On short time scales, the fluctuation-dissipation temperature is equal
to the microscopic temperature. 
As the driven system is forced to explore 
phase space further (slow relaxation), 
the correlation drops below its plateau value
and a different fluctuation-dissipation temperature
is observed, which is only weakly dependent on the bath temperature
\cite{BBK,barratkob}. 
This last time scale is strongly shear rate dependent, as is well known
from rheological experiments.

Our results were obtained on relatively short time scales, for a system 
which is usually accepted as a reasonable microscopic equivalent of molecular
or metallic glasses. 
We believe, however, that they would be most easily
applicable and experimentally testable for complex fluids, in which the glassy
state can be more easily disrupted to establish a permanent shear flow.

\acknowledgements
We acknowledge a continued collaboration and many 
useful discussions on this work  with Jorge Kurchan,
who  suggested this kind of simulation. We thank Walter Kob for a careful
reading of the manuscript.
This work was supported by the P\^ole
Scientifique de Mod\'elisation Num\'erique at ENS-Lyon
and the CDCSP at the University of Lyon.


\begin{thebibliography}{199}

\bibitem{aging_th}
J.-Ph. Bouchaud, L. F. Cugliandolo, J. Kurchan and M. M\'ezard, {\it
Spin glasses and random fields}, Ed.: A. P. Young (World
Scientific, Singapore, 1998);
T. R. Kirkpatrick and P. G. Wolynes, Phys. Rev. A {\bf 35}, 3072 (1987);
T. R. Kirkpatrick and D. Thirumalai, Phys. Rev. B {\bf 36}, 5388 (1987).

\bibitem{aging_exp}
R. L. Leheny and S. Nagel, 
Phys. Rev. B {\bf  57}, 5154 (1998);
N. E. Israeloff and T. S. Grigera,
Europhys. Lett. {\bf 43}, 308 (1998) and 
Phys. Rev. Lett. {\bf 83},5038 (1999); 
D. Bonn, J. Tanaka, G. Wegdam, H. Kellay and J. Meunier,
Europhysics Lett. {\bf 45}, 52 (1999); L. Bellon, C. Laroche and 
S. Ciliberto, Europhys. Lett. {\bf 51}, 551 (2000).

\bibitem{aging_num}
W. Kob and J.-L. Barrat,
Phys. Rev. Lett. {\bf 78}, 4581 (1997); 
W. Kob, F. Sciortino
and P. Tartaglia, Europhys. Lett. {\bf 49}, 590 (2000).


\bibitem{fdt_lj} G. Parisi, Phys. Rev. Lett. {\bf 79}, 3660 (1997);
J.-L. Barrat and W. Kob, Europhys. Lett. {\bf 46}, 637 (1999);
R. Di Leonardo, L. Angelani, G. Parisi and G. Ruocco, 
Phys. Rev. Lett. {\bf 84}, 6054 (2000).

\bibitem{leticia3} L.F. Cugliandolo, J. Kurchan and L. Peliti, Phys. 
Rev. E {\bf 55}, 3898 (1997).

\bibitem{ledou}  H. Horner, Z. Physik B {\bf 100}, 243 (1996); L. F. 
Cugliandolo, J. Kurchan, P. Le Doussal and L. Peliti,
Phys. Rev. Lett. {\bf 78}, 350 (1997).

\bibitem{BBK} L. Berthier, J.-L. Barrat and J. Kurchan, Phys. Rev. E 
{\bf 61}, 5464 (2000).

\bibitem{franz} S. Franz and M. Virasoro, J. Phys. A {\bf 33}, 891 (2000).

\bibitem{theo} T. Nieuwenhuizen, Phys. Rev. Lett. {\bf 80}, 5580 (1998).

\bibitem{peter} P. Sollich, Phys. Rev. E {\bf 58}, 738 (1998).

\bibitem{Onuki} R. Yamamoto and A. Onuki, Phys. Rev. E, {\bf 58}, 3515 (1998).

\bibitem{kob} W. Kob and H.C. Andersen,
Phys. Rev. E {\bf 53}, 4134 (1995); {\it ibid.} 
{\bf 51}, 4626 (1995); Phys. Rev.
Lett. {\bf 73}, 1376 (1994).

\bibitem{note1} Note that in previous works (\cite{kob,barratkob}),
a  different time unit, $\tau = (m_A\sigma_{AA}^2/48
\epsilon_{AA})^{1/2}$, is sometimes used when dealing with
 Lennard-Jones particles.

\bibitem{barratkob} W. Kob and J.-L. Barrat, Eur. Phys. J. B {\bf 13},
319 (2000).

\bibitem{AT87}  M.  Allen and D. Tildesley, {\it Computer simulation of
liquids} (Oxford University Press, Oxford, 1987).

\bibitem{shear_thinning}
R. G. Larson, {\it The structure and rheology of complex fluids} (Oxford
University Press, New York, 1999).


\end{thebibliography}
\end{document}